\documentclass[aps, pra, preprint, superscriptaddress, preprintnumbers, amsmath, amssymb, letterpaper, floatfix, showpacs]{revtex4-1}
\usepackage{graphicx}
\usepackage{times}
\usepackage{dcolumn}
\usepackage{array}
\usepackage{threeparttable}
\usepackage{lineno}
\newcommand{\msr}{$\mu$SR}
\newcommand{\ppg}{PrPt$_{4}$Ge$_{12}$}
\newcommand{\lpg}{LaPt$_{4}$Ge$_{12}$}
\newcommand{\plpg}{Pr$_{1-x}$La$_{x}$Pt$_{4}$Ge$_{12}$}
\newcommand{\pcpg}{Pr$_{1-x}$Ce$_{x}$Pt$_{4}$Ge$_{12}$}

\newcommand{\plpgVII}{$\mathrm{Pr}_{0.7}\mathrm{La}_{0.3}\mathrm{Pt_{4}}\mathrm{Ge_{12}}$}

\newcommand{\plpgIII}{$\mathrm{Pr}_{0.3}\mathrm{La}_{0.7}\mathrm{Pt_{4}}\mathrm{Ge_{12}}$}

\newcommand{\pos}{PrOs$_{4}$Sb$_{12}$}
\newcommand{\pors}{Pr(Os$_{1-x}$Ru$_{x}$)$_{4}$Sb$_{12}$}
\newcommand{\plos}{Pr$_{1-x}$La$_{x}$Os$_{4}$Sb$_{12}$}

\begin{document}

\preprint{ver.9-2}

\title{\boldmath Broken time-reversal symmetry in superconducting Pr$_{1-x}$La$_{x}$Pt$_{4}$Ge$_{12}$}
\author{J. Zhang}
\altaffiliation{Present address: Geballe Laboratory for Advanced Materials, Stanford University, Stanford, CA 94305, USA.}
\affiliation{State Key Laboratory of Surface Physics, Department of Physics, Fudan University, Shanghai 200433, People's Republic of China}
\author{Z. F. Ding}
\affiliation{State Key Laboratory of Surface Physics, Department of Physics, Fudan University, Shanghai 200433, People's Republic of China}
\author{K. Huang}
\altaffiliation[Present address: ]{Lawrence Livermore National Laboratory, Livermore, California 94550, USA.}
\affiliation{State Key Laboratory of Surface Physics, Department of Physics, Fudan University, Shanghai 200433, People's Republic of China}
\author{C. Tan}
\affiliation{State Key Laboratory of Surface Physics, Department of Physics, Fudan University, Shanghai 200433, People's Republic of China}
\author{A. D. Hillier}
\author{P. K. Biswas}
\affiliation{ISIS Facility, STFC Rutherford Appleton Laboratory, Harwell Science and Innovation Campus, Chilton, Didcot, Oxon, UK}
\author{D. E. MacLaughlin}
\affiliation{Department of Physics and Astronomy, University of California, Riverside, California 92521, USA}
\author{L. Shu}
\altaffiliation[Corresponding author: ]{leishu@fudan.edu.cn.}
\affiliation{State Key Laboratory of Surface Physics, Department of Physics, Fudan University, Shanghai 200433, People's Republic of China}
\affiliation{Collaborative Innovation Center of Advanced Microstructures, Nanjing 210093, People's Republic of China}

\date{\today}

\begin{abstract}
The superconducting state of the filled skutterudite alloy series~\plpg\ has been systematically studied by specific heat, zero-field muon spin relaxation (\msr), and superconducting critical field measurements. An additional inhomogeneous local magnetic field, indicative of broken time-reversal symmetry (TRS), is observed in the superconducting states of the alloys. For $x \lesssim 0.5$ the broken-TRS phase sets in below a temperature~$T_m$ distinctly lower than the superconducting transition temperature~$T_c$. For $x \gtrsim 0.5$ $T_m \approx T_c$. The local field strength decreases as $x \to 1$, where \lpg\ is characterized by conventional pairing. The lower critical field~$H_{c1}(T)$ of \ppg\ shows the onset of a second quadratic temperature region below $T_q \sim T_m$. Upper critical field~$H_{c2}(T)$ measurements suggest multiband superconductivity, and point gap nodes are consistent with the specific heat data. In \plpg\ only a single specific heat discontinuity is observed at $T_c$, in contrast to the second jump seen in \pos\ below $T_c$. These results suggest that superconductivity in \ppg\ is characterized by a complex order parameter.
\end{abstract}

\maketitle

\section{INTRODUCTION}

The study of novel superconductors with intrinsic multiple superconducting phases has broadened the understanding of the microscopic origin of unconventional superconductivity (SC)~\cite{Sigrist1991RMP, NormanScience}. A particular challenge is to clearly identify multiple SC phases in unconventional superconductors and to interpret their SC order parameter(s) (OP).

Gauge symmetry is always broken in the superconducting state. A key indication of multi-phase SC is the observation of additional broken symmetry [time-reversal symmetry (TRS), inversion symmetry] at a distinct temperature $T_m$ below the superconducting transition temperature $T_c$. However, superconductors with intrinsic multiple SC phases are extremely rare. $T_c \sim T_m$ for many superconductors with broken TRS, such as Sr$_{2}$RuO$_{4}$~\cite{Luke98}, LaNiC$_{2}$~\cite{LaNiC2}, SrPtAs~\cite{SrPtAs} and Re$_{6}$Zr~\cite{Re6Zr}. Empirically, superconductors with 4$f$ or 5$f$ electron elements are likely to have complex SC OPs~\cite{Sigrist1991RMP, Mineev1999} that can possibly lead to the emergence of multiple SC phases. For example, superconducting U$_{1-x}$Th$_{x}$Be$_{13}$~\cite{Heffner90UBe13} and UPt$_{3}$~\cite{Schemm2014TRSB} were found to be exhibit broken TRS at $T_m < T_c$, which suggests the existence of a second SC phase below $T_m$. The unusual properties of these heavy fermion (HF) superconductors have led to theories that invoke odd-parity (spin-triplet) Cooper pairing~\cite{NormanScience}.

Of particular interest in the small family of multi-phase superconductors is the puzzling SC OP of the filled-skutterudite $f$-electron compound~\pos~\cite{Maple2007POS}, which exhibits broken TRS~\cite{Aoki2003PrOsSb}\@. Several thermodynamic experiments show evidence for two SC transitions~\cite{Izawa2003pos, Falk2016PKETRSB} in \pos, although $T_m$ is not clearly below the upper $T_c = 1.85$~K in muon spin relaxation (\msr) experiments~\cite{Aoki2003PrOsSb} and it has been argued~\cite{Measson2006} that the double $T_c$ nature is not intrinsic. The SC OP of \pos\ is a consequence of crystalline-electric-field and strong spin-orbital coupling (SOC) effects~\cite{Maple2007POS}, and is complicated by the low $T_c$ and HF behavior.

The isostructural compound \ppg, with a smaller electron effective mass $m^{\ast}$, is considered to share great similarity with \pos\ in the pairing state~\cite{Gumeniuk2008PrLaPRL, Maisuradze2009Superfluid}, although no multiple SC phases have been reported. The much higher $T_c \sim 7.8$~K and non-HF state makes \ppg\ a simpler playground to study SC in Pr-based filled skutterudites. Then the key question is whether the OP in \ppg\ is complex, including the possibility of a spin-triplet state and broken TRS~\cite{Maisuradze2009Superfluid, Maisuradze2010TRSB, Kanetake2010NQR, Zhang2013multiband, Zhang2015TRSB, PFau2016La}.

The pairing symmetry of \ppg\ remains controversial. The observed broken TRS suggests an OP with either spin or orbital moments~\cite{Maisuradze2010TRSB}. Thermodynamic studies reported the presence of point-like nodes in the SC energy gap~\cite{Gumeniuk2008PrLaPRL}. Superfluid density measurements~\cite{Maisuradze2009Superfluid} suggest a non-unitary chiral $p$-wave state with gap function~$\Delta_{0}|\hat{k}_{x} \pm i\hat{k}_{y}|$, where $\Delta_0$ is the magnitude of the gap. These experimental observations motivated theoretical predictions of novel spin-nondegenerate nodal quasiparticle excitations due to strong SOC~\cite{Kozii2016Majorana}. On the other hand, the OPs of \ppg\ and the spin-singlet superconductor \lpg\ are found to be compatible~\cite{PFau2016La}, and the observation  of a Hebel-Slichter coherence peak~\cite{HSpeak1954} below $T_c$ in the $^{73}$Ge NMR relaxation rate in \ppg~\cite{Kanetake2010NQR} is evidence of weakly-coupled conventional pairing. The situation is clearly quite fluid.

In this paper, we argue that unconventional SC with a complex OP in \ppg\ is favored by our doping study of \plpg, $x = 0.1$, 0.2, 0.3, 0.5, 0.7, and 0.9, using measurements of specific heats, muon spin relaxation (\msr) in zero and longitudinal~\footnote{Parallel to the initial muon spin polarization.} field (ZF- and LF-\msr, respectively)~\cite{Brewer94, Blun99, YaDdR11}, and superconducting critical fields. Broken TRS in superconductors often results in an inhomogeneous local magnetic field~$\mathbf{B}_\mathrm{loc}$ below $T_c$, which can be readily detected by ZF-\msr\ experiments. The width~$\delta B_s$ of this field distribution is roughly $\Delta_e/\gamma_{\mu}$, where $\Delta_e$ is the contribution of the field distribution to the static Gaussian muon spin relaxation rate~$\Delta$ and $\gamma_{\mu}/2\pi$ = 135.53 MHz/T is the muon gyromagnetic ratio~\cite{Luke98, Aoki2003PrOsSb}. We find broken TRS in all samples, with a continuous decrease of $\delta B_s$ with increasing La doping. \lpg\ is a conventional superconductor~\cite{PFau2016La, Zhang2015multiband}. Intriguingly, $T_m$ is clearly lower than $T_c$ for $x \lesssim 0.5$, and $T_c - T_m$ is significant in Pr-rich alloys ($\sim$1~K in \ppg). The temperature dependence of the lower critical field~$H_{c1}(T)$ shows an anomaly at a temperature~$T_q$ $\sim$ $T_m$ in \ppg: $H_{c1} \propto T^2$ both above and below $T_q$, but with an increased coefficient below this temperature. The upper critical field~$H_{c2}(T)$ of \plpg, $x = 0$, 0.1, 0.3 and 0.7, is well described by a two-band SC model. We find evidence for point gap nodes in \plpg\ from specific heat data, but no second SC jump in the specific heat at ${\sim}T_m$ in any of the measured alloys. Our results indicate a multi-component SC OP in \ppg, which results in intrinsic multiple SC phases.

\section{SAMPLE CHARACTERIZATION AND EXPERIMENTAL METHODS}

\paragraph{Sample preparation.} Synthesis procedures for our polycrystalline \plpg\ samples ($x$ = 0, 0.1, 0.3, 0.5, 0.7, 0.8 and 0.9) were similar to those described in Ref.~\onlinecite{Gumeniuk2008PrLaPRL} and \onlinecite{huang2014synthesis}. The body-centered-cubic structure (point group~$T_{h}$, space group~$Im\mathrm{\overline{3}}$) was confirmed by Rietveld refinements of powder x-ray diffraction (XRD) patterns.
\begin{figure}[ht]
 \begin{center}
 \includegraphics[width=0.45\textwidth]{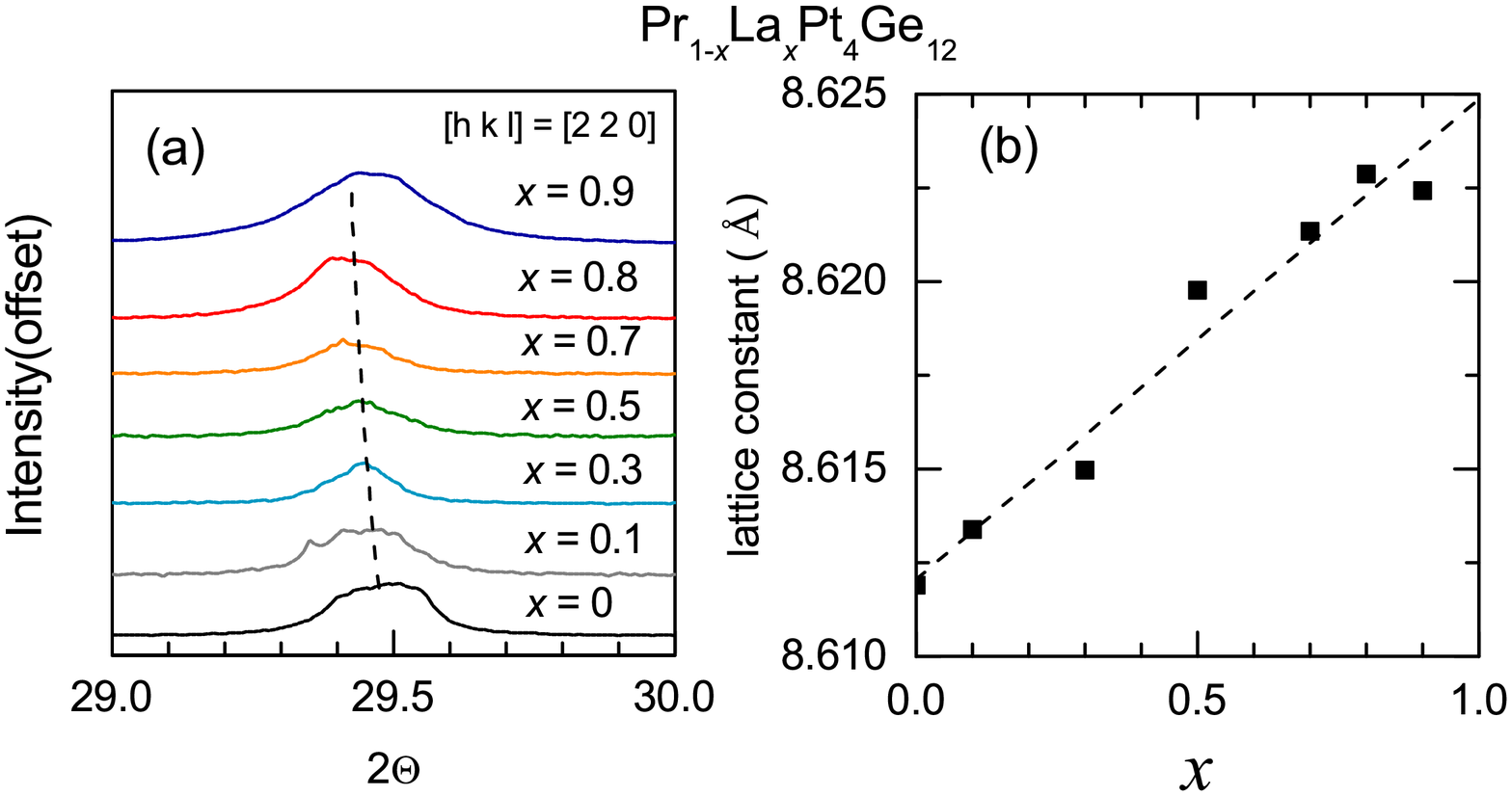}
 \caption{\label{fig:XRD} Powder X-ray diffraction results for \plpg\ ($x = 0$, 0.1, 0.3, 0.5, 0.7, 0.8 and 0.9). (a) X-ray diffraction pattern for the [h k l] = [2 2 0] peak. A gradual shift of $2\Theta$ indicates a continuous change in the lattice parameter of \plpg. (b) Lattice constant of \plpg\ calculated from XRD patterns, Vegard's law is obeyed. Dashed lines are guide to the eye.}
 \end{center}
\end{figure}
The observed isostructural linear expansion of lattice constant~$a$ with La concentration~$x$ (Fig.~\ref{fig:XRD}) is consistent with Vegard's law. No obvious Pr/La occupation defect is found.

\paragraph{Specific heat.} Measurements of the specific heat~$C_p$ of the samples were performed using a standard thermal relaxation technique in a commercial PPMS EverCool-II device (Quantum Design). Measurements on the $x$ = 0.3 and 0.7 samples were also carried out in a dilution refrigerator. The sample surfaces were polished flat to make good thermal contact. The Debye-Sommerfeld approximation~$C_p = C_e + C_\mathrm{ph} = \gamma_eT + \beta_\mathrm{ph}T^3$ was fit to the normal-state specific heat, measured in an magnetic field of 3~T over the temperature range~2--10~K\@. The Debye temperature was obtained from the relation~$\theta_{D} = \left(12\pi^4N_{A}nk_{B}/5 \beta_\mathrm{ph}\right)^{1/3}$, where $N_{A}$ is Avogadro's number, $n$ is the number of atoms per formula unit, and $k_{B}$ is the Boltzmann constant.

For \plpgVII\ an upturn in the temperature dependence of $C_p$ below $\sim$ 0.5~K was observed, due to a Schottky contribution~\cite{Aoki2002} $C_{\mathrm{Sch}}~\propto~T^{-2}$ from quadrupole-split $^{141}$Pr nuclei (data not shown). The fit gives $C_{\mathrm{Sch}}(T)/\gamma_e$ = 0.0015(3) $T^{-2}$~K, which was subtracted to obtain $C_e$ of \plpgVII. A Schottky anomaly was also reported in the parent compound \ppg\cite{Maisuradze2009Superfluid}, but is not observed in \plpgIII~down to 0.1~K\@. It is possible that there is Schottky contribution in the low-temperature $C_p$, but it is small due to diluted $^{141}$Pr nuclei and thus hard to fit and be subtracted from $C_p$\@.

\paragraph{$\mu$SR.} In time-differential \msr~\cite{Brewer94, Blun99, YaDdR11}, spin-polarized muons are implanted into the sample and decay via the weak interaction: $\mu^\pm \to e^\pm + \bar{\nu}_\mu(\nu_\mu) + \nu_e(\bar{\nu}_e)$. The decay positron (electron) count rate asymmetry~$A_\mu(t)$ in a given direction is proportional to the muon ensemble spin polarization~$P_\mu(t)$ in that direction, and yields information on static and dynamic local fields at the muon sites. Positive muons ($\mu^{+}$), which stop at interstitial sites in solids, are generally used because they sample local magnetism better than negative muons, which are tightly bound to nuclei.

\msr\ experiments were carried out on samples with $x$ = 0.3, 0.5, 0.7, 0.8 and 0.9 using the MUSR spectrometer at the ISIS Neutron and Muon Facility, Rutherford Appleton Laboratory, Chilton, UK, over the temperature range~0.3--12~K\@. The ambient magnetic field was actively compensated to better than 1~$\mu$T\@. The ZF-\msr\ data for the parent compound~\ppg\ discussed in this paper are from Ref.~\onlinecite{Zhang2015TRSB}.

The observed ZF and LF asymmetry time spectra~$A_\mu(t) = a_0P_\mu(t)$, where $a_0$ is the initial asymmetry, are well described by the exponentially-damped Gaussian Kubo-Toyabe polarization function~\cite{KT}
\begin{equation} \label{eq:edgkt}
P_{\mu}(t) = e^{-\lambda t}G_F^\mathrm{KT}(\Delta, t),
\end{equation}
where $F = \text{ZF}$ or LF, $\Delta/\gamma_\mu$ is the rms width of a Gaussian distribution of static local fields, and the damping rate~$\lambda$ is often interpreted as a dynamic relaxation rate due to thermal fluctuations of the muon local field~\footnote{But not always; see Sec.~\ref{sec:lfmusr}.}. Here
\begin{equation} \label{eq:zfgkt}
G_\mathrm{ZF}^\mathrm{KT}(\Delta, t) = \frac{1}{3} + \frac{2}{3} \left(1 - \Delta^{2}t^{2}) \exp (- \textstyle{\frac{1}{2}} \Delta^{2}t^{2}\right)
\end{equation}
and
\begin{eqnarray} \label{eq:lfgkt}
G_\mathrm{LF}^\mathrm{KT}(\Delta_\mathrm{LF}, t) & = & 1 - \frac{2\Delta_\mathrm{LF}^2}{\omega_L^2}\left[1 - \exp\left(-{\textstyle\frac{1}{2}}\Delta_\mathrm{LF}^2t^2\right)\cos(\omega_Lt)\right] \nonumber \\
& + & \frac{2\Delta_\mathrm{LF}^4}{\omega_L^3} \int_0^t \exp\left(-{\textstyle\frac{1}{2}}\Delta_\mathrm{LF}^2\tau^2\right)\sin(\omega_L\tau)\, d\tau,
\end{eqnarray}
where $\omega_L = \gamma_\mu H_L$ is the muon Zeeman frequency in the longitudinal field~$H_L$. Equations~(\ref{eq:edgkt})--(\ref{eq:lfgkt}) have previously been used to analyze data from \pos\ and \ppg\ and their alloys~\cite{Aoki2003PrOsSb, Maisuradze2010TRSB, Shu2011TRSB, Zhang2015TRSB}.

\paragraph{Magnetic susceptibility.} Susceptibility measurements were carried out using a commercial vibrating sample magnetometer (VSM) (Quantum Design) down to 2~K\@. Lower critical fields $H_{c1}$ were determined from the field dependence of the superconducting magnetization~$M_s(H,T)$ measured after cooling in ``zero'' field (less than $\sim$0.3 mT)\@. $H_{c1}$ was defined typically as the field value where the magnetization deviates from the field shielding initial slope. Upper critical fields~$H_{c2}$ were taken as the onset of $M_s(H,T)$ in field-cooled experiments. Results are discussed in Sec.~\ref{sec:Hc1Hc2}.

\section{RESULTS} \label{sec:results}

The lattice constant~$a$, superconducting transition temperatures~$T_c^{C_p}$ from the specific heat, Sommerfeld coefficients~$\gamma_e$, Debye temperatures~$\theta_{D}$, and the relative specific heat discontinuity~$\Delta {C_p}/\gamma_eT_c$ at $T_c$ are summarized in Table~\ref{tb:TableI}, including data for the two end compounds ($x = 0$ and 1) from Refs.~\onlinecite{Gumeniuk2008PrLaPRL} and \onlinecite{Maisuradze2009Superfluid}.
\begin{table*}[ht]
\caption{\label{tb:TableI} Properties of \plpg: lattice constant~$a$, superconducting transition temperature~ $T_c$ determined from specific heat~$C_p(T)$, Sommerfeld specific heat coefficient~$\gamma_e$, relative discontinuity~$\Delta C_p/\gamma_eT_c$ in $C_p(T)$ at $T_c$, Debye temperature~$\theta_D$, and coefficient~$b$ in Eq.~(\ref{eq:empirical}) (see text).}
\begin{ruledtabular}
\begin{tabular}{d|cccccccc}
$x$ & $a$ (\AA) & $T_c$ (K) & $\gamma_e$ ($\mathrm{mJ/mol~K^2}$) & $\Delta C_p/\gamma_eT_c$ & $\theta_D$ (K) & $b$ \\
\hline
 0.0\, \footnotemark[1] \footnotemark[2] & 8.6111(6) & 7.91  & 87.1 & 1.56 & 198 & 1.2(1) \\
 0.1 & 8.6134(1) & 7.85 & 76.3 (6) & 1.56 & 201 & -- \\
 0.3 & 8.6150(2) & 7.87 & 67.9 (5) & 1.67 & 204 & 1.2(2) \\
 0.5 & 8.6198(5) & 7.97 & 70.5 (7) & 1.55 & 203 & 1.3(2) \\
 0.7 & 8.6213(4) & 8.02 & 68.8 (7) & 1.54 & 216 & 1.3(2) \\
 0.8 & 8.6229(4) & 8.07 & 67.2 (6) & 1.53 & 208 & 2.1(7) \\
 0.9\, \footnotemark[3] & 8.6224(3) & 8.22 & 64.6 (6) & 1.53 & 208 & 2.0(-) \\
 1.0\, \footnotemark[1] & 8.6235(3) & 8.27 & 75.8 & 1.49 & 209 & -- \\
\end{tabular}
\end{ruledtabular}
\footnotetext[1]{Data from Ref.~\onlinecite{Gumeniuk2008PrLaPRL}.}
\footnotetext[2]{Data from Ref.~\onlinecite{Maisuradze2009Superfluid}.}
\footnotetext[3]{For $x$ = 0.9, $\Delta_e(T)$ is scattered, and a fit using Eq.~(\ref{eq:empirical}) with $b$ free does not converge well. A fit with fixed $b = 2.0$, which gives reasonable fit quality, is used to obtain $\Delta_e(0)$.}
\end{table*}
 The quantity~$b$ is the coefficient in Eq.~(\ref{eq:empirical}) of Sec.~\ref{sec:zfmusr}, where $\mu$SR relaxation data are analyzed assuming the contribution~$\Delta_e(T)$ to the static relaxation rate due to broken TRS exhibits a BCS gap-like temperature dependence.

\subsection{Specific Heat}

\begin{figure}[ht]
 \includegraphics[width=0.45\textwidth]{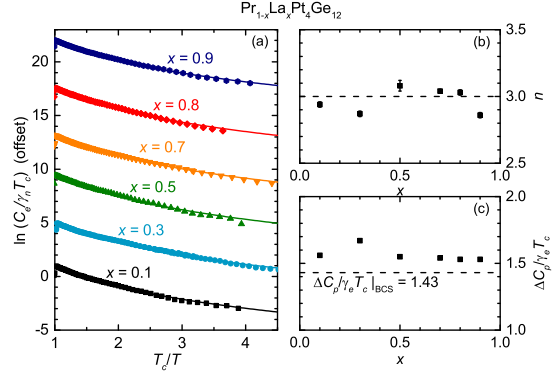}
 \caption{\label{fig:spht1} (a)~Temperature dependence of the electronic specific heat~$C_e$ plotted as $\ln(C_e/\gamma_eT_c)$ vs. $T_c/T$, with offsets for clarity. Curves represent power law fit to $C_e$ vs. $T$. (b)~La doping dependence of $n$, where $n$ is the power of the $C_e \propto T^n$ fit below $T_c$. (c)~La doping dependence of $\Delta C_{p}/\gamma_e T_c$. For all $x$, the values of $\Delta C_{p}/\gamma_e T_c$ are larger than the BCS value of 1.43.}
\end{figure}

\begin{figure}[ht]
 \includegraphics[width=0.45\textwidth]{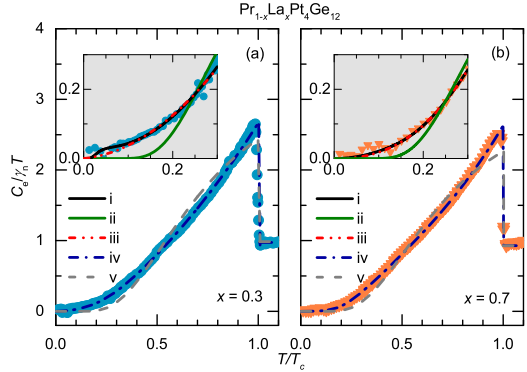}
 \caption{\label{fig:spht2} (a) and (b)~Temperature dependence of $C_e/T$ in zero applied field for $x = 0.3$ and 0.7, respectively. Insets: fits at low $T$\@. Curves:~(i)~Low-$T$ weighted sum of asymptotic BCS term (see text) and $T^{3}$. (ii)~Low-$T$ single asymptotic BCS term~\cite{aymptoticBCS}. (iii)~Low-$T$ weighted sum of two power-law terms $T^{2}$ and $T^{3}$. Fits up to $T_c$: (iv)~Weighted sum of power-law term and BCS empirical term. (v)~Single BCS empirical term.}
\end{figure}

 The temperature dependencies of the electronic specific heat~$C_e(T)$ for \plpg, $x = 0.1$, 0.3, 0.5, 0.7, 0.8 and 0.9, are shown in Fig.~\ref{fig:spht1} and Fig.~\ref{fig:spht2}.
There is only one obvious SC specific heat jump for all La concentrations, so that macroscopic separation of \ppg\ and \lpg\ SC phases due to sample inhomogeneity is unlikely. The nonlinearity of the $T_c/T$ dependence of $\ln$($C_e$/$\gamma_e$$T_c$) for all measured samples [Fig.~\ref{fig:spht1}(a)] suggests structure in the superconducting gap~\cite{Sigrist1991RMP, huang2014synthesis}. A simple power-law fit to $C_e$ vs. $T$ ($C_e \propto T^n$) can give heuristic estimation of the gap symmetry. The values of the power $n$ are displayed in Fig.~\ref{fig:spht1}(b). A $\sim T^3$ dependent $C_e$ suggests point gap nodes in $x \neq 1$ alloys. Also, the values of $\Delta C_{p}/\gamma_e T_c$ for all measured samples are larger than the BCS value of 1.43 (Fig.~\ref{fig:spht1}(c)), indicating unconventional SC.

The temperature dependencies of the specific heats for $x = 0.3$ and 0.7 alloys down to $T/T_c \approx 0.02$ are displayed in Figs.~\ref{fig:spht2}(a) and \ref{fig:spht2}(b). For $T < T_c$, $C_e(T)$ can be best described by the two-component ``$\alpha$ model''~\cite{Bouquet2001alphamodel} with a weighting factor $f$:
\begin{equation}
\label{eq:alpha}
C_e = f C_{e1} + (1-f) C_{e2}\,.
\end{equation}
This phenomenological model has been used to describe multi-band superconductors such as MgB$_{2}$~\cite{MgB2}, and Eq.~(\ref{eq:alpha}) with both $C_{e1}$ and $C_{e2}$ based on the BCS theory with different gaps was used to characterize \lpg~\cite{PFau2016La}. In \plpgVII\ and \plpgIII, however, $C_e(T)$ can be best fit using $C_{e1} \propto e^{-\Delta_1/k_{B}T}$ and $C_{e2} \propto T^n$, which are empirical expressions for a full gap and a gap with nodes, respectively~\cite{Sigrist1991RMP}. This parameterization results in $f = 0.17$, $n = 2.91$, and $\Delta_1/k_{B}T_c = 1.33$ for \plpgVII; and $f = 0.26$, $n = 3.1$, and $\Delta_1/k_{B}T_c = 1.13$ for \plpgIII. We are not able obtain acceptable fits using other forms for $C_{e1}$ and $C_{e2}$. The decrease in $f$ with increasing $x$ suggests the gradual disappearance of the point-node gap as $x$ increases, and for $x = 1$, where TRS is fully restored, there is no power-law term ($f = 1$)~\cite{Zhang2015multiband}.

At low temperatures a single fully-gapped scenario does not yield the measured $C_e(T)$ of either \plpgVII\ or \plpgIII; neither the asymptotic form~\cite{aymptoticBCS}
\begin{equation}
\label{eq:asymptoticBCS}
\lim\limits_{T\rightarrow 0}\frac{C_e}{\gamma_e T}= 3.15\left(\frac{\Delta_{0}}{1.76k_{B}T}\right)^{5/2}\!\exp\left(-\frac{\Delta_{0}}{k_{B}T}\right)
\end{equation}
nor a simple exponential fit the data. The best fits to the data at low temperatures ($0.1~\mathrm{K} < T < 0.3T_c$) are to Eq.~(\ref{eq:alpha}) with the low-temperature asymptotic form for $C_{e1}(T)$ and a power law for $C_{e2}(T)$ with $n = 3$. Fits to weighted sums of $T^{2}$ and $T^{3}$ result in $f$~$\rightarrow$~0 for $T^{2}$, precluding the possibility of line nodes in the gap. The goodness of fit for single power-law $T^{3}$ fits is worse than for multi-component fits.

Thus our results suggest that \plpg\ alloys are multiband superconductors, with one of the gap functions characterized by point-like nodes.

Our specific heat measurements in \plpg\ are in good agreement with previous work~\cite{Gumeniuk2008PrLaPRL, Maisuradze2009Superfluid, PFau2016La}, except that reported values of the Sommerfeld coefficient~$\gamma_e$ for $x = 0$ and 1 (Table~\ref{tb:TableI}) are somewhat higher than extrapolations from the alloys. Differences between reported values have been noted previously, and may be due to measurements at different fields and temperatures~\cite{huang2014synthesis}. There is no evidence for sample differences based on comparison with published results~\cite{Gumeniuk2008PrLaPRL, huang2014synthesis}; in particular, we do not observe the dramatic decrease in $\gamma_e$ found in Pr$_{0.5}$Pt$_4$Ge$_{12}$~\cite{Venkatesh2014PPG} and attributed to Pr vacancies.

\subsection{Muon Spin Relaxation} \label{sec:musr}

\subsubsection{Zero-Field \msr} \label{sec:zfmusr}

\begin{figure}[ht]
 \includegraphics[width=0.45\textwidth]{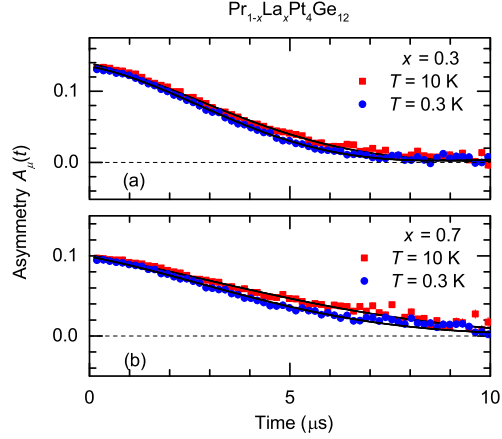}
 \caption{\label{fig:Asy} Representative muon asymmetry spectra~$A_{\mu}(t)$ for \plpg, $x = 0.3$ and 0.7, measured above and below $T_c$ (squares and circles, respectively). A background signal has been subtracted from the data. Solid curves: fits using the exponentially damped Kubo-Toyabe function [Eqs.~(\ref{eq:edgkt}) and (\ref{eq:zfgkt})].}
\end{figure}

Figure~\ref{fig:Asy} shows representative \msr\ asymmetry spectra measured in ZF for \plpg, $x = 0.3$ and 0.7, above and below $T_c$.
A non-relaxing background signal originating from muons that miss the sample and stop in the silver sample holder has been subtracted from the data. As in \ppg~\cite{Maisuradze2010TRSB, Zhang2015TRSB}, no early-time oscillations or fast relaxation is observed, indicating the absence of strong static magnetism (with or without long-range order). The small additional relaxation below $T_c$ indicates the emergence of a distribution of weak spontaneous local fields~\cite{KT}.

\begin{figure}[ht]
 \includegraphics[width=0.45\textwidth]{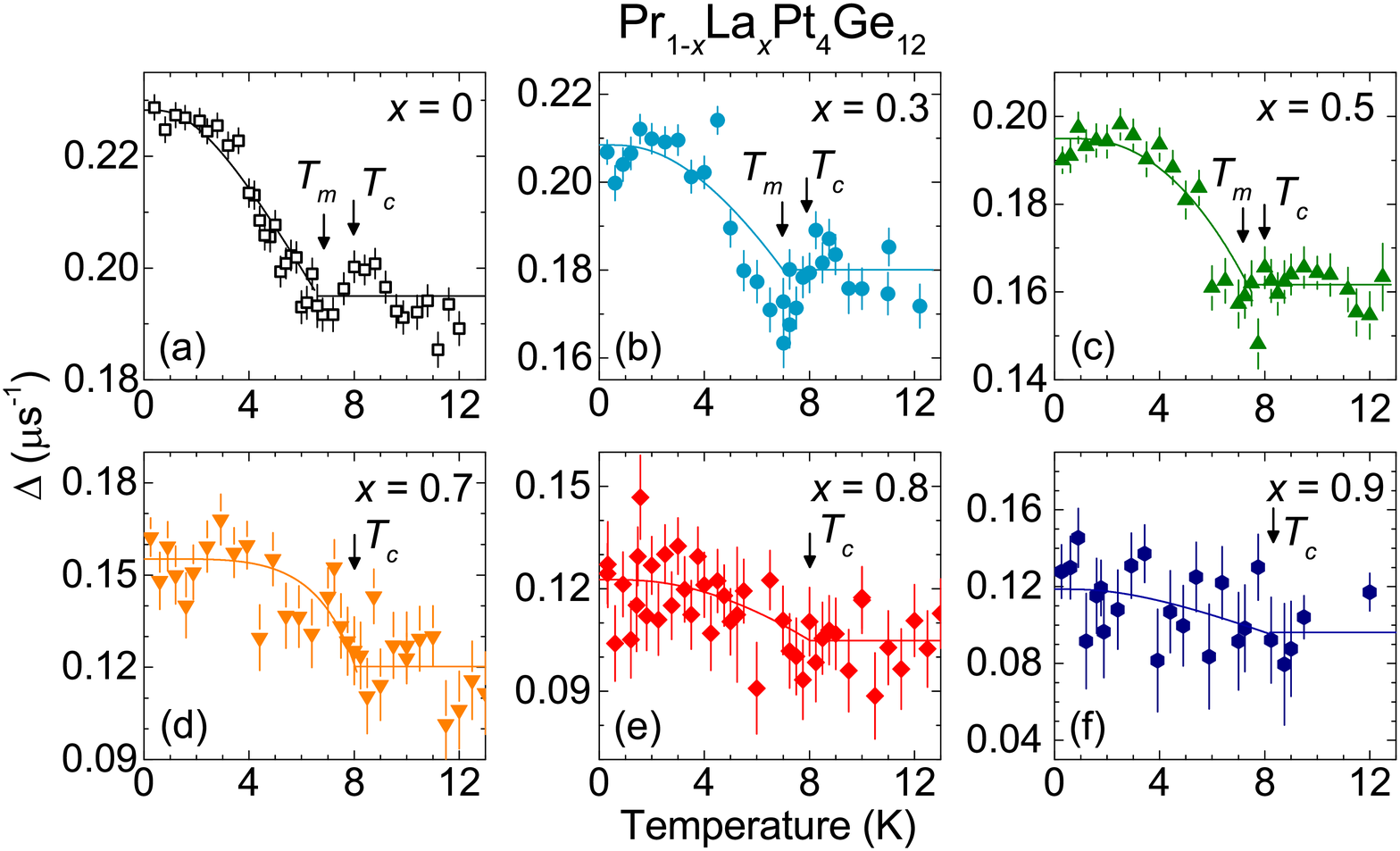}
 \caption{ \label{fig:Delta} (a)--(f): Temperature dependence of the ZF KT static Gaussian relaxation rate~$\Delta$ in \plpg, $x = 0$, 0.3, 0.5, 0.7, 0.8, and 0.9 samples, respectively. Curves: fits of Eqs.~(\ref{eq:Deltae}) and (\ref{eq:empirical}) assuming a BCS-like temperature dependence. Arrows: $T_c$ from specific heat, $T_m$ from onset of broken TRS.}
\end{figure}

Figure~\ref{fig:Delta} gives the temperature dependence of the Gaussian relaxation rate~$\Delta$ in \plpg, $x = 0$, 0.3, 0.5, 0.7, 0.8 and 0.9, obtained from fits of Eqs.~(\ref{eq:edgkt}) and (\ref{eq:zfgkt}) to the data.
In the normal state $\Delta$ is temperature independent, as expected for muon depolarization by dipolar fields from quasistatic nuclear moments~\cite{KT}. The enhancement of $\Delta(T)$ with decreasing temperature below $T_c$ is due to the onset of static local fields, and is strong evidence for broken TRS~\cite{Sigrist1991RMP, MacKenzieSr2RuO4RMP}\@. With increasing La content the enhancement decreases, and disappears at $x = 1$ where TRS is fully restored~\cite{Maisuradze2010TRSB}. Our results for $x = 0$ agree with those of Ref.~\onlinecite{Maisuradze2010TRSB}.

This increase is of electronic origin, so that the electronic and random nuclear contributions~($\Delta_e$ and $\Delta_n$, respectively) are uncorrelated and add in quadrature~\cite{Aoki2003PrOsSb}:
\begin{equation} \label{eq:Deltae}
\Delta(T) = \left[\Delta_n^{2} + \Delta_e^{2}(T)\right]^{1/2}.
\end{equation}
The curves in Fig.~\ref{fig:Delta} are fits of Eq.~(\ref{eq:Deltae}) to the data using the approximate empirical expression~\cite{Gross1986BCS}
\begin{equation} \label{eq:empirical}
\Delta_e(T) = \Delta_e(0) \tanh \left(b\sqrt{\frac{T_m}{T} - 1}\right) \quad (T < T_m),
\end{equation}
assuming that $\Delta_e$ has the temperature dependence of a BCS-like OP but the transition temperature is $T_m$ rather than $T_c$. In \plpg\ the coefficient~$b$ (Table \ref{tb:TableI}) is somewhat smaller than the isotropic BCS value of 1.74 in the weak coupling limit.

The dip followed by an increase in $\Delta(T)$ below $T_c$ (Fig.~\ref{fig:Delta}) was also noted in Ref.~\onlinecite{Maisuradze2010TRSB}, and its reproducibility suggests that it is not an instrumental effect. Screening of magnetic impurity dipolar fields was proposed as its origin~\cite{Maisuradze2010TRSB}. Later work~\cite{Zhang2015TRSB} indicated that this mechanism is too weak to produce the effect, however, and its origin is not understood.

\begin{figure}[ht]
 \includegraphics[clip=,width=0.45\textwidth]{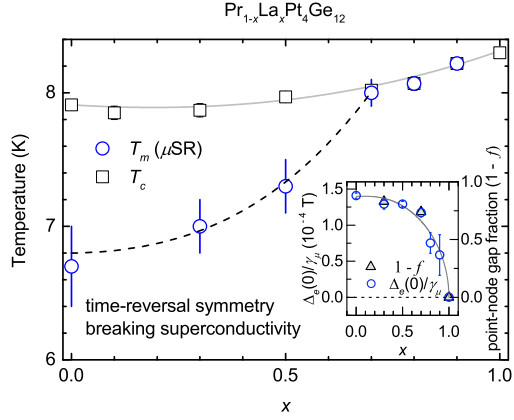}
 \caption{\label{fig:PD} Temperature-La concentration~$x$ phase diagram for \plpg. Squares: $T_c$ from specific heat data. Circles: onset temperature~$T_m$ of broken TRS\@. Inset: circles: $x$ dependence of local spontaneous magnetic field distribution width~$\Delta_e(0)/\gamma_{\mu}$ Triangles: point-like node gap fraction $1 - f$ from specific heat results [Eq.~(\ref{eq:alpha})]. Curves are guides to the eye.}
\end{figure}

Figure~\ref{fig:PD} gives the $T$-$x$ phase diagram for the \plpg\ alloy system.
Intriguingly, $T_m < T_c$ for $x \lesssim 0.5$, as previously reported in \ppg~\cite{Maisuradze2010TRSB} where it is particularly obvious. The zero-temperature width~$\Delta_e(0)/\gamma_{\mu}$ of the broken-TRS field distribution~\footnote{Note that this field is an effect of broken TRS and not its cause; in a pure superconductor with orbital broken TRS the field would be Meissner screened~\cite{Sigrist1991RMP}.} is shown in the inset of Fig.~\ref{fig:PD}. It is interesting to note that $\Delta_e(0)/\gamma_{\mu}$ and the pointlike node gap fraction from specific heat results (inset of Fig.~\ref{fig:PD} have the same $x$ dependence, which is highly nonlinear compared to that in the \pors, \plos\ or \pcpg\ alloy systems~\cite{Shu2011TRSB, Zhang2015TRSB}. We thus speculate that the origin of the broken TRS in \plpg\ is strongly correlated with the anisotropic gap structure. Noting that $T_m < T_c$ for $x \lesssim 0.5$, we also speculate that interaction between neighboring $^{141}$Pr$^{3+}$ ions contributes to the difference between $T_m$ and $T_c$.

\begin{figure}[ht]
 \includegraphics[width=0.45\textwidth]{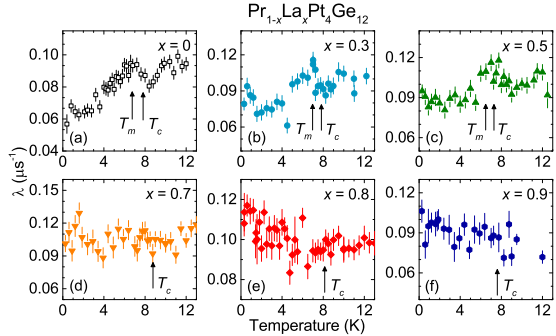}
 \caption{\label{fig:Lambda} (a)--(f): Temperature dependence of the zero-field exponential damping rate~$\lambda$ in \plpg, $x = 0$, 0.3, 0.5, 0.7, 0.8 and 0.9.}
\end{figure}

Figure~\ref{fig:Lambda} shows the ZF exponential damping rate~$\lambda(T)$ in \plpg, $x = 0$, 0.3, 0.5, 0.7, 0.8 and 0.9.
In \ppg\ and Pr-rich alloys there is a distinct maximum in $\lambda(T)$ below $T_c$, followed by a decrease to a nonzero value as $T \to 0$. In the normal state above $T_c$ $\lambda(T)$ increases with increasing temperature. The temperature-dependent features become smaller with increasing $x$, and for $x = 0.7$ the rate is essentially temperature-independent. For $x = 0.8$ $\lambda(T)$ increases with decreasing temperature below ${\sim}5~\text{K} \approx 0.6T_c$, and for $x = 0.9$ there are not enough normal-state data to determine a temperature dependence. The La concentration dependence of the zero-field $\lambda(T)$ in \plpg\ (Fig.~\ref{fig:Lambda}) is similar to that observed in \pcpg, which is discussed in some detail by Zhang \textit{et~al.}~\cite{Zhang2015TRSB}.

The maximum in $\lambda(T)$ is accompanied by a dip in $\Delta(T)$ at ${\sim}T_c$ (Fig.~\ref{fig:Delta}). This might be an effect of statistical correlation between the two parameters, but the dip in $\Delta$ is still present when $\lambda(T)$ is fixed in (poorer-quality) fits of Eqs.~(\ref{eq:edgkt}) and (\ref{eq:zfgkt}) to $P_\mu(t)$.

Mechanisms for $\lambda(T)$ are discussed further in Sec.~\ref{sec:dynamic}.

\subsubsection{\boldmath Longitudinal-Field \msr} \label{sec:lfmusr}

In ZF the dynamic and static field contributions to the muon spin relaxation rate are hard to disentangle experimentally. The usual procedure for doing this is measurement of $A_\mu(t)$ in a longitudinal field~$\mathbf{H}_L$ parallel to the initial muon spin polarization~$\mathbf P_\mathrm\mu(0)$. For $H_L$ much greater than static local fields at muon sites ($\omega_L \gg \Delta$), the resultant static field is nearly parallel to $\mathbf P_\mathrm\mu(0)$, so that there is little static muon relaxation (precession and dephasing). Then $\mathbf P_\mathrm\mu(t)$ is said to be ``decoupled'' from the static local fields, and any remaining relaxation for high $H_L$ is dynamic in origin, due to thermal fluctuation of the local fields.

A field dependence of the dynamic relaxation rate is also expected, due to the proportionality of $\lambda$ to the fluctuation noise power at $\omega_L$; in general this is reduced at high frequencies. If the fluctuations are characterized by a correlation time~$\tau_c$, $\lambda(H_L)$ is given by the Redfield equation~\cite{[] [{, Chap.~5.}] Slichter}
\begin{equation} \label{eq:Redfield}
 \lambda_\mathrm{LF}(H_L) = \frac{\delta\omega_\mathrm{rms}^2 \tau_c}{1 +(\omega_L\tau_c)^2}
\end{equation}
in the motionally-narrowed limit~$\delta\omega_\mathrm{rms}\tau_c \ll 1$, where $\delta\omega_{\mathrm{rms}} = \gamma_\mu\langle \delta B_{\mathrm{loc}}^{\mathrm{2}}(t) \rangle^{1/2}$ is the rms fluctuating local field amplitude in frequency units. Thus $\lambda_\mathrm{LF}(H_L)$ decreases with increasing~$H_L$ for $\omega_L\tau_c \gtrsim 1$. This field dependence can be distinguished from decoupling if $1/\tau_c$ is sufficiently larger than the static rate~$\Delta$.

\begin{figure}[ht]
 \includegraphics[clip=, width=0.45\textwidth]{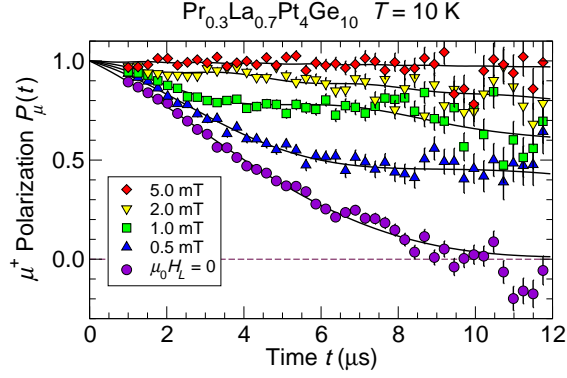}
 \caption{\label{fig:10Kpol} Longitudinal field $H_{\mathrm{L}}$ dependence of $\mu^+$ polarization time spectra from \plpgIII, $T = \text{10 K} > T_c$. Curves: fits of Eqs.~(\ref{eq:edgkt}) and (\ref{eq:lfgkt}) to the spectra.}
\end{figure}

LF-\msr\ experiments were carried out in the normal state ($T = 10$~K) of \plpgIII. Muon spin polarization time spectra~$P_\mu(t)$, obtained from the asymmetry~$A_\mu(t)$ by subtracting the signal from the cold finger and normalizing, are shown in Fig.~\ref{fig:10Kpol}.
The field dependence of $P_\mu(t)$ shows characteristics of both decoupling and field-dependent dynamic relaxation: the small-amplitude oscillation with frequency~$\omega_L$ at intermediate fields is a feature of decoupling~\cite{KT}, but decoupling alone would not account for the nonzero field dependence of the overall relaxation rate at intermediate fields.

\begin{figure}[ht]
 \includegraphics[clip=, width=0.45\textwidth]{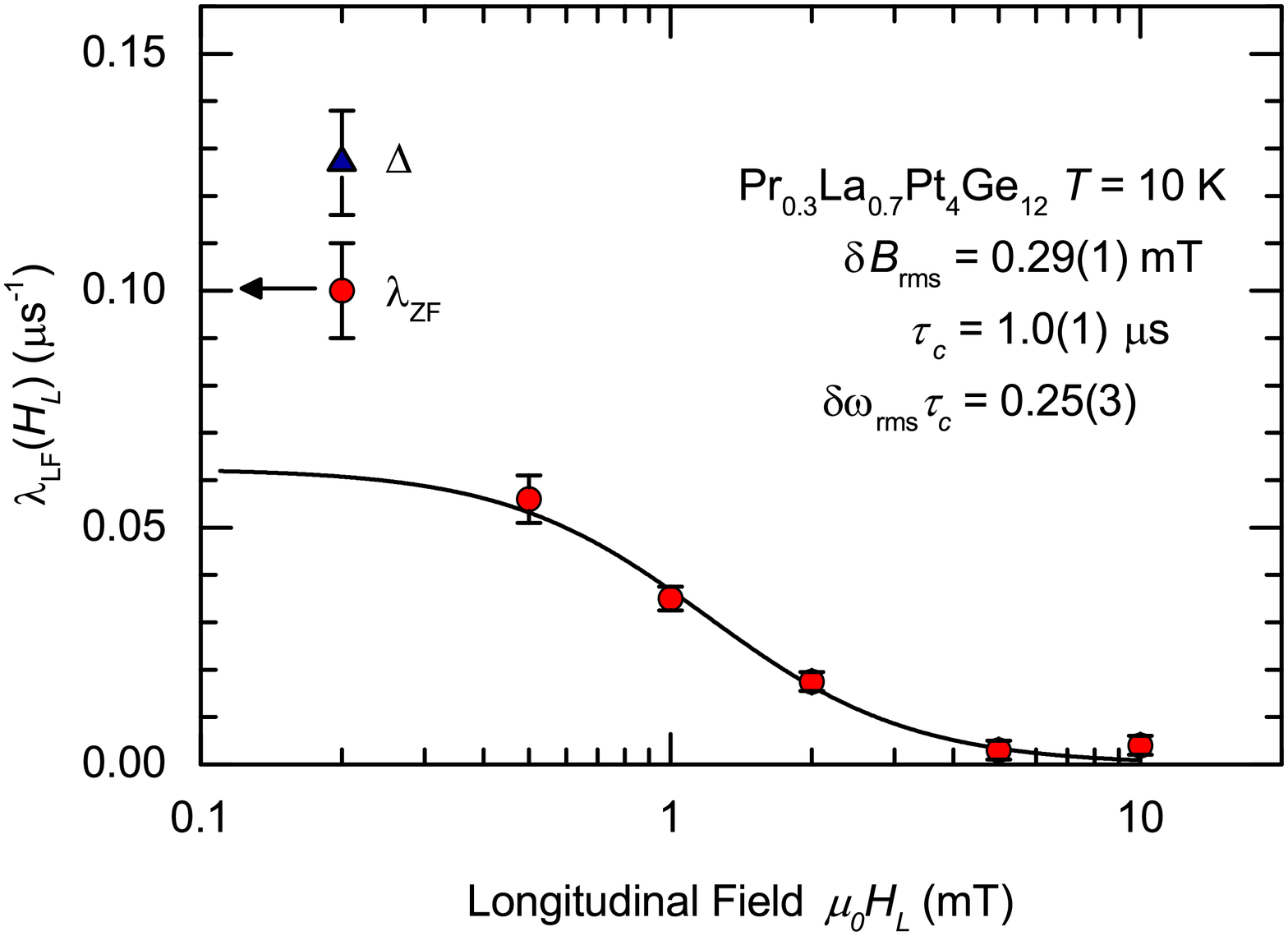}
 \caption{\label{fig:LF} Field dependence of the relaxation rate~$\lambda_\mathrm{LF}$ in \plpgIII, $T = 10$~K\@. Curve: fit of Eq.~(\ref{eq:Redfield}) to the data for $\mu_0H_L \geqslant 0.5$~mT.}
\end{figure}

These data have been fit using Eqs.~(\ref{eq:edgkt}) and (\ref{eq:lfgkt}). The field dependence of $\lambda_\mathrm{LF}$ for \plpgIII\ at 10~K is shown in Fig.~\ref{fig:LF}.
For $\mu_0H_L \geqslant 0.5$~mT a good fit to Eq.~(\ref{eq:Redfield}) is obtained, which yields $\tau_c = 1.0(1)~\mu$s and $\delta B_{\mathrm{rms}} = 0.29(1)$~mT\@. The motional-narrowing criterion~$\delta\omega_\mathrm{rms}\tau_c \ll 1$ is more or less satisfied [$\delta\omega_\mathrm{rms}\tau_c = 0.25(3)$], but the local field fluctuations are close to ``adiabatic.'' The same behavior is observed in \pos~\cite{Aoki2003PrOsSb}, with similar parameter values.

Values of the static rate~$\Delta$ and the ZF damping rate~$\lambda_\mathrm{ZF}$ are also shown in Fig.~\ref{fig:LF}. The relation~$\Delta\tau_c \ll 1$, necessary to separate decoupling and dynamic-rate field dependences, is also more or less satisfied [$\Delta\tau_c = 0.14(1)$]. The fit value~$\lambda_\mathrm{LF}(H_L{\to}0) = 0.062(2)~\mu\text{s}^{-1}$ is less than $\lambda_\mathrm{ZF}$, suggesting a static contribution to the latter. This would indicate a so-called Voigtian static field distribution, i.e., the convolution of Gaussian and Lorentzian distributions~\cite{CrCy97}. An early report of a Voigtian distribution in \ppg~\cite{Maisuradze2010TRSB} was not reproduced in later studies~\cite{Zhang2015TRSB}.

\subsubsection{Mechanism for Dynamic Muon Melaxation?} \label{sec:dynamic}

There are two candidate sources of fluctuating $\mu^+$ local fields in \plpg: Pr$^{3+}$ $4f$ fluctuations, and nuclear magnetism. There is no Pr$^{3+}$ local moment, since the Pr$^{3+}$ non-Kramers crystal-field ground state is nonmagnetic. Although $4f$/conduction-band mixing is likely in these alloys, the $\mu^+$ contact interaction with conduction or $4f$ electrons is normally weak and relaxation rates are too slow to be observed in the \msr\ time window. This can be seen from an estimation of the conduction-electron hyperfine coupling constant $\omega_\mathrm{hf}$ (or hyperfine field~$H_\mathrm{hf} = \omega_\mathrm{hf}/\gamma_\mu$) needed to yield the observed $\mu^+$ relaxation rate, since $\lambda \approx \omega_\mathrm{hf}^{2} \tau_c$, and for conduction-electron relaxation $\tau_c$ is at the most of the order of $\hbar/k_{B}T_{\mathrm{F}} \sim 10^{-13}$~s, where $T_{\mathrm{F}} \sim 100$~K is the ``Fermi temperature'' from the specific heat of \ppg~\cite{huang2014synthesis} and associated with a putative $4f$ band. With the observed $\lambda \sim 0.1~\mu\text{s}^{-1}$ this yields $\mu_0H_\mathrm{hf} \sim 1$~T, which is is an order of magnitude larger than typical muon/conduction-electron or muon/$4f$ hyperfine fields in metals~\cite{Schenck85}.

In Pr-rich alloys $^{141}$Pr nuclei are dominant. In \ppg, as in \pcpg~\cite{Zhang2015TRSB}, the $\mu^+$ $\lambda(T)$ resembles the $^{73}$Ge NMR relaxation rate~\cite{Kanetake2010NQR}, and is similarly reminiscent of Hebel-Slichter relaxation~\cite{HSpeak1954} in fully-gapped superconductors. This suggests $^{141}$Pr spin fluctuations with correlation times~$\tau_c$ slower than the motionally-narrowed limit, in which case $\lambda \sim 1/\tau_c$.

The Hebel-Slichter rate vanishes exponentially as $T \to 0$, whereas in \ppg\ and its alloys the muon spin relaxation remains nonzero at low temperatures. We note, however, that mechanisms for fluctuating nuclear magnetism are of two kinds: spin-lattice ($T_1$) relaxation, due to interactions with the electronic environment, and spin-spin ($T_2$) relaxation, due to interactions between nuclei~\footnote{Ref.~\cite{Slichter}, Chap.~3.}. The latter are not expected to be temperature dependent at ordinary temperatures. Contributions of both mechanisms to nuclear spin fluctuations would be consistent with the observed behavior.

There is a problem with this scenario in Pr-rich alloys, however: the $^{141}$Pr dipolar fields should be fluctuating fully (i.e., have no static component), in which case the "dynamic" K-T relaxation function~\cite{KT} is appropriate. But as in \pcpg~\cite{Zhang2015TRSB} fits to this function are not as good as fits using Eqs.~(\ref{eq:edgkt}) and (\ref{eq:zfgkt}). This is perhaps less of a difficulty for La-rich alloys, where $^{139}$La nuclear spin fluctuations are expected to be much slower and could account for the static relaxation. The reduction with increasing $x$ of the temperature dependence of $\lambda$ (which is nearly $T$-independent in \plpgIII) is consistent with this possibility, but the increased $\lambda(T)$ below $T_c$ for $x = 0.8$ is not understood. The situation remains unclear, and more work needs to be done.

\subsection{Critical Fields} \label{sec:Hc1Hc2}

\begin{figure}[ht]
 \includegraphics[width=0.45\textwidth]{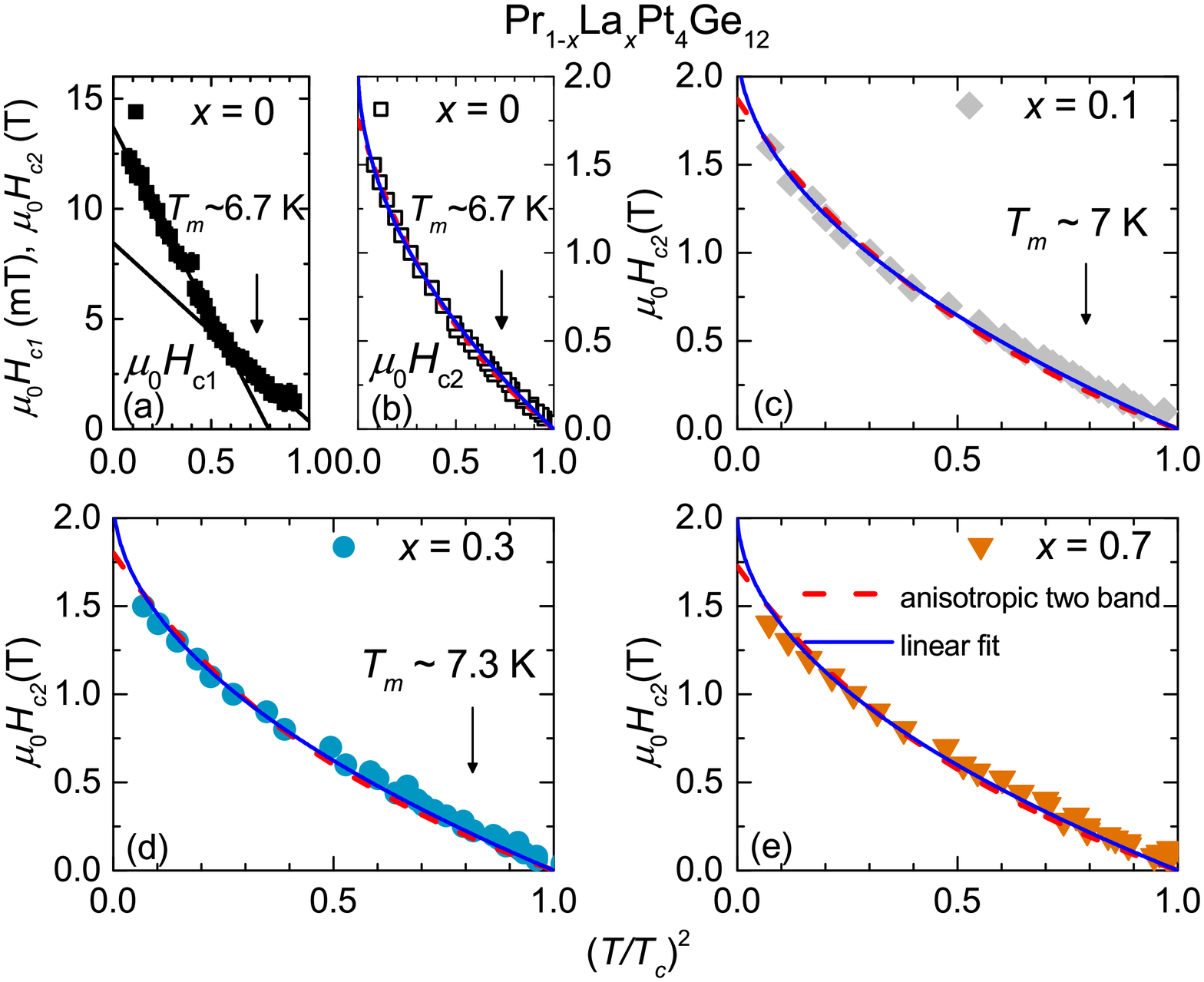}
 \caption{\label{fig:CriticalField} Critical-field study of \plpg\ ($x$~=~1, 0.9, 0.7, and 0.3). Arrows: values of $T_m$.
 (a) Lower critical field $H_{c1}(T)$ of $x$ = 0. Two quadratic temperature regions are observed with lower $T^{2}$ onset temperature $T_q$~=~6~K\@. (b)-(e) Upper critical field $H_{c2}(T)$ of $x$ = 0, 0.1, 0.3 and 0.7, respectively. Dashed curves are fit to Eq.~\ref{eq:alpha2}. Solid curves are a linear $T/T_c$ fit to $H_{c2}(T)$, assuming \plpg\ is a two-band superconductor in the dirty limit.}
\end{figure}

The origin of the difference between $T_m$ and $T_c$ has been further investigated by lower and upper critical field measurements. Results are shown in Fig.~\ref{fig:CriticalField}.
As shown in Fig.~\ref{fig:CriticalField}(a), we observe the possible existence of a second quadratic temperature dependence region below $T_q$ $\sim$ 6~K in $H_{c1}(T)$ of \ppg. Within errors $T_q$ is close to $T_m$, but is slightly higher in a similar $H_{c1}(T)$ study~\cite{ChandraHCPPG}. In contrast, $H_{c2}(T)$ does not exhibit enhancement below $T_m$, but rather a smooth increase with decreasing temperature for all the measured \plpg\ as shown in Figs.~\ref{fig:CriticalField}(b)-(e).

The origins of $H_{c1}(T)$ and $H_{c2}(T)$ are different. $H_{c1}(T)$ might provide a heuristic description of the gap symmetry, since in the London limit (coherence length $\xi$ much smaller than the penetration depth $\Lambda$)~\footnote{This criterion is satisfied, since $\xi (0)$ = [$\mathrm{\Phi_{o}}$/$2\pi H_{c2}(0)$]$^{1/2}$ = 13 nm taking $H_{c2}(0)$ = 1.75 T (from our $H_{c2}(T)$ analysis of the anisotropic two-band model) and $\Lambda$ = 114 (4) nm from \msr~\cite{Maisuradze2009Superfluid}.} $H_{c1}(T) / H_{c1} (0)$ is proportional to the superfluid density $\rho_s$~$\propto$~$n_s/m^{*}$~\cite{Tinkham75}. For a fully gapped conventional superconductor such as pure niobium, $H_{c1}(T)$/$H_{c1}(0)$ = 1 - ($T$/$T_c$)$^{2}$~\cite{FRENCHNb}. The occurrence of the second quadratic temperature region thus indicates gap structure, possibly multiband structure or gap anisotropy.

Considering the onset of broken TRS at $T_m$, the extra enhancement of $H_{c1}(T)$ below $T_q$ could imply the occurrence of a distinct magnetic phase below $T_c$ as in U$_{1-x}$Th$_{x}$Be$_{13}$~\cite{Heffner90UBe13}. Intriguingly, Ref.~\onlinecite{Maisuradze2009Superfluid} showed that $\rho_s/\rho_{0}(T/T_c)$ measured at 35 mT slightly deviates from other field results at $T/T_c$ $\sim$ 0.75 (close to $T_q$). This deviation was attributed to larger vortex disorder at fields close to $H_{c1}$), but an anomaly in $\rho_s/\rho_{0}(T/T_c)$ (35 mT) might be relevant. It should be mentioned, however, that \msr\ experiments found no obvious anomaly at $T_m$ in this quantity, measured at various fields between 75 mT and 640 mT~\cite{Maisuradze2009Superfluid}. The upturn in $H_{c1}(T)$ could also be due to a flux pinning effect, which can play a significant role in determining $H_{c1}(T)$ on decreasing temperature. A similar situation occurs in isostructural \pos, where enhancement of $H_{c1}(T)$ below 0.6~K was found~\cite{CichorekPOS} but there is no anomaly at this temperature in specific heat or \msr\ results~\cite{DougPOS2010}.

As shown in Figs.~\ref{fig:CriticalField}(b)-(e), the temperature dependence of the upper critical field $H_{c2}$ of \plpg\ provides evidence of multiband SC. The curves in these panels are fits to
\begin{equation}
\label{eq:alpha2}
H_{c2}(T)/H_{c2}(0)=\frac{1-(T/T_c)^{2}}{1+(T/T_c)^{2}},
\end{equation}
which is an empirical description appropriate to anisotropic two-band superconductors~\cite{CHANGJAN2011Hc2}. Fit parameters~$\mu_0H_{c2}(0)$ = 1.75(1) T, 1.87(3) T, 1.80(3) T and 1.73(3) T are found for $x$ = 0, 0.1, 0.3 and 0.7, respectively. Moreover, a linear fit of $T/T_c$ dependence of $H_{c2}(T)$ also falls on most of the data. Linear fits~$H_{c2}(T)/H_{c2}(0) = 1 - T/T_c$ give $\mu_{0}H_{c2}(0)$ = 2.07 (2) T, 2.20 (0.04) T, 2.13 (2) T and 2.04 (3) T for $x$ = 0, 0.1, 0.3 and 0.7, respectively. A quasi-linear $T$ dependence is also observed in the multi-band superconductor MgB$_{2}$~\cite{Cristina2001}, where SC can be explained by the dirty two-gap quasi-classical Usadel equations~\cite{Gurevich2003}. This suggests that SC in \plpg\ could be described by the Ginzburg-Landau theory in the dirty limit with nonmagnetic intraband and interband impurity scattering. However, results on single crystalline \ppg\ estimated the mean free path $l_{\mathrm{mfp}}$ = 103 nm using $\gamma_e$ from specific heat and $\rho$ from resistivity measurements~\cite{Zhang2013multiband}. This is much larger than the coherence length $\xi$(0), and suggests that \ppg\ is in the local and clean limit~\cite{Zhang2013multiband}. It is therefore evidence that the linearity of $H_{c2}(T)$ is not of dirty-limit two-band origin. It should be mentioned that in polycrystalline \ppg, the mean free path $l_{\mathrm{mfp}}$ is estimated to be close to $\xi$(0) using the free electron theory, and $l_{\mathrm{mfp}}$ is much smaller than the Pippard coherence length $\xi_p$ $\sim$ 360 nm~\cite{Venkatesh2014PPG}. We conclude that more work is required to understand this situation.

A linear fit to $H_{c2}(T)$ data near $T_c$ (from $T_c$ to 0.8~K below) gives initial slopes~$|$d$H_{c2}$/d$T$$|$$_{T_c}$ = 1.80(11) T~K$^{-1}$, 1.74(17) T~K$^{-1}$, 2.15(24) T~K$^{-1}$ and 1.64(17) T~K$^{-1}$ for $x$ = 0, 0.1, 0.3 and 0.7, respectively. In a one-band superconductor, the Werthamer-Helfand-Hohenberg (WHH) theory~\cite{WHH} predicts an orbital limiting field $H_{c2}^{\mathrm{orb}}$(0) = 0.73 $|$d$H_{c2}$/d$T$$|$$_{T_c}$ $T_c$ in the clean limit, and $H_{c2}^{\mathrm{orb}}$(0) = 0.69 $|$d$H_{c2}$/d$T$$|$$_{T_c}$ $T_c$ in the dirty limit. In either case, $H_{c2}^{\mathrm{orb}}$(0) is smaller than the expected weak coupling Pauli limit field $H_{c2}^{\mathrm{P}}$(0) = 1.84$T_c$ = 14.6 T assuming a spin singlet state. However, both $H_{c2}^{\mathrm{orb}}$(0) and $H_{c2}^{\mathrm{P}}$(0) are much larger than $H_{c2}(0)$. We thus speculate that the single-band Ginzburg-Landau theory does not describe SC well in \ppg; the SC OP is more complex.

\section{DISCUSSION}

\subsection{Multiband Superconductivity}

We have shown that in \plpg\ upper critical fields $H_{c2}(T)$ are consistent with two SC gaps. The two-band $\alpha$ model is capable of describing the specific heat results. We found that the larger-gap band has point nodes and the smaller-gap band is weakly coupled. The small fraction~$f$ of the BCS gap for $x$ = 0.3 and 0.7 suggests that SC of \plpg\ is dominated by the larger band with point nodes. Values of $\Delta {C_p}/\gamma_eT_c$ for \plpg\ are all larger than the BCS value of 1.43, indicating that \ppg\ is in the strong coupling limit. This conforms to the theoretical constraints that the larger gap is strongly coupled and the smaller gap is weakly coupled~\cite{Kresin1990}. Pr-based filled skutterudites, including \pos~\cite{Lei2009MBSC, SeyfarthPOS, HillPOSPRS}, PrRu$_{4}$As$_{12}$~\cite{NamikiPrRuAs} and PrRu$_{4}$Sb$_{12}$~\cite{TakedaPrRuSb, HillPOSPRS}, might all be multiband superconductors.

Maisuradze \textit{et~al.}~\cite{Maisuradze2009Superfluid} concluded that in polycrystalline \ppg\ the gap functions~$|$$\mathrm{\Delta_{0}}$sin$\theta$$|$ and $\mathrm{\Delta_{0}}$(1-sin$^{4}$$\theta$cos$^{4}$$\theta$), both with point nodes, can best describe the gap symmetry. These are also candidate gap functions for \pos~\cite{ChiaPOS, MakiPOS}. However, a study of single-crystalline \ppg~\cite{Zhang2013multiband} reports two isotropic BCS gaps. While our $H_{c2}(T)$ and $C_e(T)$ results suggest multiband SC, point gap nodes in one band are favored by $C_e(T)$ results, corresponding well to the \msr\ study~\cite{Maisuradze2009Superfluid}. The discrepancy in $C_e(T)$ between polycrystalline and single-crystalline \ppg\ was previously attributed to the inaccurate subtraction of the nuclear Schottky anomaly at low temperature~\cite{Zhang2013multiband}. Although accurate subtraction is difficult, this only affects the analysis of $C_e(T)$ at low temperatures, and fits over the entire temperature range $< T_c$ yield evidence for point gap nodes. Furthermore, the Schottky anomaly is reduced and not observed in $x$ $>$ $0.5$ alloys, whereas $C_e(T)$ $\propto$ $T^{3.1}$ is still found in \plpgIII. High quality single crystalline \plpg\ samples will be required to clarify the origin of this discrepancy.

It should be mentioned that we did not observe any obvious specific heat jump around $T_{m}$ in both $x = 0$ and $x = 0.3$ with a $C_e (T)$ measurement $T$ step $\Delta T$ = 0.025 K. Moreover, $C_e (T)$ anomaly at $T_m$ was not reported in several previous work, including the single crystal study by Zhang $et~al.$~\cite{Zhang2013multiband}.

\subsection{Broken Time Reversal Symmetry}

In general, the violation of TRS naturally occurs in bulk superconductors with non-unitary states~\cite{Sigrist1991RMP, SigristTRSB2000}. Broken TRS can emerge at temperatures differing from $T_c$, such as in the $E_{2u}$ odd-parity triplet state candidate UPt$_{3}$~\cite{Schemm2014TRSB}. The appealing claim of spin-triplet chiral $p$ wave state in \ppg\ was first anticipated from superfluid density results~\cite{Maisuradze2009Superfluid}. To date experimental confirmations of other implications are still lacking.

A singlet pairing state cannot be ruled out, due to the continuous variation of $\delta B_s(x)$ as well as the smooth and small change of $T_c(x)$ in \plpg. These results suggest a continuous variation of OP symmetry between \lpg\ and \ppg~\cite{Maisuradze2010TRSB, PFau2016La}, where \lpg\ is characterized by a spin-singlet state.

Broken TRS tends to be stabilized by the forming of a subdominant gap~\cite{SigristTRSB2000}. Thus if a spin-singlet state is present in \ppg, a straightforward explanation of broken TRS, point gap nodes together with a BCS gap, and $T_m < T_c$ might be chirality in a superconducting state with an $s$-wave component~\cite{Kozii2016Majorana}, e.g., the $d + is$ state~\cite{SigristTRSB2000}. Then supercurrents form around non-magnetic imperfections, resulting in local fields and the enhancement of $\Delta(T)$ below $T_m$~\cite{Choi1989, Maisuradze2010TRSB}. A subdominant $s$-wave component in the SC OP of \ppg\ results in the two-consecutive SC phase feature~\cite{SigristTRSB2000, SigristTRSB20002}.

It was observed~\cite{ChandraHCPPG} that in \ppg\ the critical current density $I_c$ decreases exponentially with increasing applied field and decreases linearly with increasing temperature. The absence of an anomaly in $I_c(H)$ around $T_m$ indicates no macroscopic separation of multiple SC phases~\cite{Venkatesh2014PPG}, but is not evidence against the claim of multiple components in the SC OP.

The model of Koga, Matsumoto, and Shiba (KMS)~\cite{KMS}, which proposed several $T_{h}$ symmetric irreducible representations for isostructural \pos, suggests that in this compound SC with broken TRS is driven by crystal-field excitonic Cooper-pairing~\cite{Shu2011TRSB}. With increasing La concentration the weakened Pr-Pr intersite interaction results in a crossover between SC ground states with broken and preserved TRS due to less excitonic dispersion, resulting in a monotonic decrease in $\delta B_s(x)$ in \plos~\cite{Shu2011TRSB}.

To some extent \plpg\ resembles \plos, since both end compounds are superconductors and they both exhibit a continuous decrease in $\delta B_s$ on increasing $x$. Considering the small variation of $T_c$ and the smooth evolution of $\delta B_s(x)$ in \plpg, the KMS model might be applicable to describe the SC OP in this alloy series. Sergienko and Curnoe~\cite{Sergienko2004theory} also proposed a singlet-state model for \pos, based on the assumptions of point gap nodes as well as equivalent OP between the TRS breaking SC phase and the TRS-preserved one. These assumptions are compatible with our results in \ppg. In this case the absence of multiple SC specific heat jumps in tetrahedral \plpg\ is a reflection of orbital components in the OP, and the broken TRS originates from a complex orbitally degenerate OP in the spin-singlet state within the strong SOC limit~\cite{Sergienko2004theory, Maisuradze2010TRSB}.

It should be mentioned that orbital degrees of freedom further complicate the origin of the SC excitation gap. It was found that even-parity superconductors with broken TRS have multiband OP~\cite{Sigrist1991RMP, BogoliubovFS}. These superconductors are characteristic of two-dimensional Fermi surfaces, with nodes in the gap replaced by Bogoliubov quasiparticles. \ppg\ could be such candidate material considering the broken TRS state with multiband feature. Moreover, it was recently proposed that three-dimensional chiral superconductors with strong SOC and odd-parity state can be the host of gapless Majorana fermions~\cite{Kozii2016Majorana}. \pos\ was proposed as a candidate for such a system, due to the existence of point gap nodes, broken TRS, and a possible spin-triplet state. Isostructural \ppg\ was also proposed to be a candidate material. The two-band feature is not evidence against a non-unitary state~\cite{Kozii2016Majorana}. We thus note here that the origin of the multiband feature with nodal gap structure and broken TRS in \ppg\ remains an unsolved problem.

\section{CONCLUDING REMARKS}

We report specific heat, ZF- and LF-\msr, and critical field studies of well-characterized polycrystalline superconducting \plpg, to investigate the SC OP of the parent compound \ppg. There is no obvious SC pair breaking effect in \plpg, since $T_c(x)$ varies only slightly. We find that \plpg\ alloys exhibit broken TRS, with a signature local field distribution width $\delta B_s$ below $T_c$ that continuously decreases with increasing $x$, disappearing for $x\sim 1$. This behavior is a reflection of a crossover between broken and preserved TRS SC across the alloy series, similar to \plos. The most intriguing feature is that the onset temperatures~$T_m$ for broken TRS in \plpg\ ($x$ $\lesssim$ 0.5) are below $T_c$, as is most obvious in \ppg. Furthermore, $H_{c1}(T)$ exhibits a second quadratic temperature region below $T_q \approx T_m \sim$ 6~K\@. These results are likely due to the intrinsic two-consecutive-SC-phase nature of \ppg. Critical fields~$H_{c2}(T)$ and the specific heat~$C_e(T)$ in \plpg\ can be well described by a two-band model, with $C_e(T)$ suggesting a dominant point-node gap structure in the pairing symmetry. However, there are no obvious multiple SC specific heat jumps at $T_m$. Based on these results, we conclude that \ppg\ is likely to be a multiband superconductor characterized by a complex multi-component SC OP with orbitally degenerate representations. Our results motivate further studies of SC excitation gap symmetry in \ppg\ and isostructural compounds, including \pos. We conclude that \plpg\ is a unique candidate for further study of unconventional SC with intrinsic multiple phases.

\bigskip \begin{acknowledgments} We wish to thank M. B. Maple, G. M. Zhang, O. O. Ber\-nal, and P.-C. Ho for fruitful discussions. We are grateful to the ISIS Cryogenics Group for their valuable help during the \msr\ experiments. The research performed in this study was supported by the National Key Research and Development Program of China (Nos.~2017YFA0303104 and 2016YFA0300503), and the National Natural Science Foundation of China under Grant nos.~11474060 and 11774061. Research at UC Riverside (UCR) was supported by the UCR Academic Senate.
\end{acknowledgments}

%

\end{document}